\documentclass[12pt,oneside]{article}
\usepackage{amsfonts,amssymb,graphicx}
\newcommand{\Av}{{\rm Av}}
\newcommand{\p}{\mathbb P}

\def\be{\begin{equation}}
\def\ee{\end{equation}}
\def\bea{\begin{eqnarray}}
\def\eea{\end{eqnarray}}
\def\c {\bar{c}}
\def \cc {{\cal C}}
\def\<{\langle}
\def\>{\rangle}
\def\~{\tilde}
\def\s{\sigma}

\def\L{\Lambda}
\def\Lp{\Lambda^\prime}

\def\b{\beta}
\def\g{\gamma}

\def\t{\tau}

\def\ds{\displaystyle}

\newcommand{\qed}{\hfill \ensuremath{\Box}}
\newcommand{\R}{\Bbb R}

\newcommand{\Z}{\Bbb Z}

\newcommand{\av}[1]{\mbox{{\rm Av}}\left(#1\right)}

\newtheorem{theorem}{Theorem}

\newtheorem{lemma}{Lemma}

\newcommand{\beq}{\begin{eqnarray}}
\newcommand{\eeq}{\end{eqnarray}}
\newcommand{\avh}{\mbox{Av}_\xi}
\newcommand{\avk}{\mbox{Av}_\eta}
\newcommand{\avmk}{\mbox{Av}_{-\eta}}
\newcommand{\avhk}{\mbox{Av}_{\xi,\eta}}

\newcommand{\hl}{H_\Lambda}

\newcommand{\hlp}{H_{{\Lambda}^\prime}}
\newcommand{\hcl}{H_{{\Lambda \setminus {\Lambda}^\prime}}}
\newcommand{\hllp}{H_{{\Lambda, {\Lambda}^\prime}}}
\newcommand{\bgh}{{\omega}_{\xi}}
\newcommand{\bgk}{{\omega}_{\eta}}
\newcommand{\bgmk}{{\omega}_{-\eta}}
\newcommand{\bgmh}{{\omega}_{-\xi}}

\newcommand{\nsubset}{\subset \hspace{-0.18cm} \setminus \hspace{0.1cm}}
\newcommand{\bll}{{\cal X}_{\Lambda,{\Lambda}^\prime}}
\newcommand{\X}{{\cal X}}

\newcommand{\si}{\ensuremath{\sigma}}
\newcommand{\sigmap}{\sigma^\prime}
\newcommand{\gammap}{\gamma^\prime}
\newcommand{\taup}{\tau^\prime}
\newcommand{\CC}{\mathcal{C}}
\newcommand{\bp}{{b^\prime}}
\newcommand{\sint}{\sin(t)}
\newcommand{\sins}{\sin(s)}
\newcommand{\cost}{\cos(t)}
\newcommand{\coss}{\cos(s)}

\newcommand{\sinsq}{\sin^2(s)}
\newcommand{\costq}{\cos^2(t)}
\newcommand{\cossq}{\cos^2(s)}
\newcommand{\dd}[2]{\delta_{#1,#2}}
\begin{document}
\begin{center}
\vspace{1truecm}
{\bf\sc\Large  interaction-flip identities in spin glasses}\\
\vspace{1cm}
{Pierluigi Contucci$^{\dagger}$, Cristian Giardin\`a$^{\ddagger}$, Claudio Giberti$^{\star}$}\\
\vspace{.5cm}
{\small $\dagger$ Dipartimento di Matematica} \\
{\small Universit\`a di Bologna,40127 Bologna, Italy}\\
{\small {e-mail: {\em contucci@dm.unibo.it}}}\\
\vspace{.5cm}
{\small $\ddagger$ Eindhoven University of Technology \& EURANDOM}\\
{\small P.O. Box 513 - 5600 MB Eindhoven, The Netherland}\\
{\small {e-mail: {\em c.giardina@tue.nl}}}\\
\vspace{.5cm}
{\small $\star$ Dipartimento di Scienze e Metodi dell'Ingegneria } \\
{\small Universit\`a di Modena e Reggio Emilia, 42100 Reggio Emilia, Italy}\\
{\small {e-mail: {\em claudio.giberti@unimore.it}}}\\
\vskip 1truecm
\end{center}
\vskip 1truecm
\begin{abstract}\noindent
We study the properties of fluctuation for the free energies and internal energies
of two spin glass systems that differ for having some set of interactions flipped. We show that their difference
has a variance that grows like the volume of the flipped region. Using a new interpolation
method, which extends to the entire circle the standard interpolation technique, we show by integration
by parts that the bound imply new overlap identities for the equilibrium state. As a side result
the case of the non-interacting random field is analyzed and the triviality of its overlap distribution proved.
\end{abstract}
\newpage\noindent

\section{Introduction and results}

In this paper we investigate a new method to obtain overlap
identities for the spin glass models. The strategy we use is the
exploitation of a bound on the fluctuations of a quantity that
compares a system with some Gaussian disorder with the system at a
flipped ($J\to -J$) disorder. While the disordered averages are
symmetric by interaction flip due to the symmetry of the
distribution, the difference among them is an interesting random
variable whose variance can be shown to grow at most like the volume
(for extensive quantities).

The identities are then deduced using some form of integration by parts in the same
perspective in which they appear from stochastic stability \cite{AC} or the Ghirlanda Guerra
method \cite{GG} in the mean field case or, more recently, for short range finite
dimensional models \cite{CGi, CGi2} (see also \cite{T,B}).

The interest of obtaining further information from the method of the
identities lies on the fact that they provide a constraint for the
overlap moments (or their distribution) and have the potential to
reduce its degrees of freedom toward, possibly, a {\it mean field}
structure like it is expected to have the Sherrington Kirkpatrick
model.


More specifically the results of this paper consist of overlap
identities for the quenched state which interpolate between a
Gaussian spin glass and the system where the couplings in a
subvolume (possibly coinciding with the whole volume) have been
flipped. The interpolation is obtained by extending to the whole
circle the Guerra Toninelli interpolation \cite{GT}. The bounds are
derived from the concentration properties of the difference of the
free energy per particle in the two settings, original and flipped.

As an example, one may consider the result which is stated in
\cite{NS} (and quoted there as proved by Aizenman and Fisher) for
the difference $\Delta F$ between the free energy of the
Edwards-Anderson model on a $d$-dimensional lattice with linear size
$L$ and a volume $L^d$ when going from periodic to antiperiodic
boundary conditions on the hyperplane which is orthogonal to (say)
the $x$-direction. The mentioned property is a bound for the
variance of this quantity which grows no more than the volume of the
hyperplane. Such an upper bound is equivalent to a bound for the
stiffness exponent $\theta \le (d-1)/2$ \cite{SY,BM,FH} (See also the 
discussion of that exponent in \cite{vE}). Although
that bound is not expected to be saturated we prove here that it
implies an identity for the equilibrium quantities. When expressed
in terms of spin variables some of the overlap identities that we
find generalize the structure of truncated correlation function that
appear in \cite{Te} whose behaviour in the volume is related to the
low temperature phase properties of the model. Consequences of our
bound can also be seen at the level of the difference of internal
energies. This second set of identities contains as a particular
case some of the Ghirlanda-Guerra identities.

A quite interesting result, from the mathematical physics perspective, is
provided by the analysis of the identities for the random field model without interaction.
We show here that the new set of identities that we derive (and explicitly test) when
considered together with the Ghirlanda Guerra ones provide a simple
proof of triviality of the model i.e. the proof that the overlap is a non fluctuating
quantity. We plan to apply the same method to the investigation of the random
field model with ferromagnetic interactions.

The plan of the paper is the following. In the next section
we define the setting of Gaussian spin glasses that we  consider.
Then in section 3 we  prove a lemma for the first two moments
of the difference of free energies. This is obtained by studying
a suitable interpolation on the circle for the linear combination
of two independent Hamiltonians. Section 4 contains the proof
of the concentration of measure results. The main results are
given in section 5 and 6, where the new overlap identities
are stated. Finally in section 7 we study the case of the
random field model and shows how to derive the triviality of the model
without making use of the explicit solution.

\section{Definitions}
\label{def}

We consider a disordered model of Ising configurations
$\s_n=\pm 1$, $n\in \Lambda\subset {\cal L}$ for some subset
$\Lambda$ (volume $|\Lambda|$) of a lattice ${\cal L}$. We denote
by $\Sigma_\Lambda$ the set of all $\s=\{\s_n\}_{n\in \Lambda}$, and
$|\Sigma_\Lambda|=2^{|\Lambda|}$. In the sequel the
following definitions will be used.

\begin{enumerate}

\item {\it Hamiltonian}.\\ For every $\Lambda\subset {\cal L}$ let
$\{H_\Lambda(\sigma)\}_{\s\in\Sigma_N}$
be a family of
$2^{|\Lambda|}$ {\em translation invariant (in distribution)
Gaussian} random variables defined, in analogy with \cite{RU}, according to
the  general representation
\be
H_{\Lambda}(\s) \; = \; - \sum_{X\subset \Lambda} J_X\s_X
\label{hami}
\ee
where
\be
\s_X=\prod_{i\, \in X}\s_i \; ,
\ee
($\s_\emptyset=0$) and the $J$'s are independent Gaussian variables with
mean
\be\label{mean_disorder}
{\rm Av}(J_X) = 0 \; ,
\ee
and variance
\be\label{var_disorder}
{\rm Av}(J_X^2) = \Delta^2_X  \; .
\ee
Given any subset $\Lp\subseteq \L$, we also write
\be
\hl(\s)=\hlp(\s)+\hcl(\s)
\ee
where
\be\label{hamiloc}
\hlp(\s)= - \sum_{X\subset \Lp} J_X\s_X,\quad \hcl(\s)= - \sum_{X\subset \L \atop X \nsubset \Lp} J_X\s_X\;,
\ee
and
\be
\hllp(\s)=-\hlp(\s)+\hcl(\s)
\ee
will denote the Hamiltonian with the $J$ couplings inside the region $\Lp$ that have been flipped.
\item {\it Average and Covariance matrix}.\\
The Hamiltonian $H_{\Lambda}(\s)$ has covariance matrix
\begin{eqnarray}\label{cov-matr}
\label{cc}
{\cal C}_\Lambda (\s,\tau) \; &:= &\;
\av{H_\Lambda(\s)H_\Lambda (\tau)}
\nonumber\\
& = & \; \sum_{X\subset\Lambda}\Delta^2_X\s_X\t_X\, .
\end{eqnarray}
By the Schwarz inequality
\be\label{sw}
|{\cal C}_\Lambda (\s,\t)| \; \le \; \sqrt{{\cal C}_\Lambda
(\s,\s)}\sqrt{{\cal C}_\Lambda (\t,\t)} \; = \;
\sum_{X\subset\Lambda}\Delta^2_X
\ee
for all $\s$ and
$\t$.
\item {\it Thermodynamic Stability}.\\
The Hamiltonian (\ref{hami}) is thermodynamically stable if there exists
a constant $\c$ such that
\begin{eqnarray}
\label{thst}
\sup_{\Lambda\subset {\cal L}}
\frac{1}{|\Lambda|}\sum_{X\subset\Lambda}\Delta^2_X
\; & \le & \; \c \; < \; \infty\;.
\end{eqnarray}
Thanks to the relation (\ref{sw}) a thermodynamically stable model fulfills the bound
\begin{eqnarray}
\label{pippo}
{\cal C}_\Lambda (\s,\t) \; & \le & \; \c \, |\Lambda|
\end{eqnarray}
and has an order $1$ normalized covariance
\begin{eqnarray}\label{norm_covar_matrix}
c_{\Lambda}(\s,\t) \; & : = & \; \frac{1}{|\Lambda|}{\cal C}_\Lambda (\s,\t)\;.
\end{eqnarray}
\item {\it Random partition function}.
\be\label{rpf}
{\cal Z}_\Lambda(\beta) \; := \; \sum_{\s  \in \,\Sigma_\Lambda}
e^{-\beta{H}_\Lambda(\s)}\equiv \sum_{\s  \in \,\Sigma_\Lambda}e^{-\b\hlp(\s)-\b\hcl(\s)}
\; ,
\ee
\be\label{rpfflip}
{\cal Z}_{\Lambda,\Lp}(\beta) \; := \; \sum_{\s  \in \,\Sigma_\Lambda}
e^{-\beta\hllp(\s)}\equiv\sum_{\s  \in \,\Sigma_\Lambda}
e^{\b\hlp(\s)-\b\hcl(\s)}
\; .
\ee
\item {\it Random free energy/pressure}.
\be\label{rfe}
-\beta {\cal F}_\Lambda(\beta) \; := \; {\cal P}_\Lambda(\beta) \; := \; \ln {\cal Z}_\Lambda(\beta)
\; ,
\ee
\be\label{rfeflip}
-\beta {\cal F}_{\Lambda,\Lp}(\beta) \; := \; {\cal P}_{\Lambda,\Lp}(\beta) \; := \; \ln {\cal Z}_{\Lambda,\Lp}(\beta)
\; .
\ee

\item {\it Random internal energy}.
\be\label{rie}
{\cal U}_\L(\beta) \; := \; \frac{\sum_{\s  \in \,\Sigma_\Lambda}
H_{\Lambda}(\s)e^{-\beta{H}_\Lambda(\s)}}{\sum_{\s  \in \,\Sigma_\Lambda}
e^{-\beta{H}_\Lambda(\s)}}
\; ,
\ee
\be\label{rie_flip}
{\cal U}_{\L,\Lp}(\beta) \; := \; \frac{\sum_{\s  \in \,\Sigma_\Lambda}
\hllp(\s)e^{-\beta\hllp(\s)}}{\sum_{\s  \in \,\Sigma_\Lambda}
e^{-\beta\hllp (\s)}}
\; .
\ee
\item {\it Quenched free energy/pressure}.
\be
-\beta F_{\Lambda}(\beta) \; := \; P_{\Lambda}(\beta) \; := \; \av{ {\cal P}_{\Lambda}(\beta) }\; .
\ee
\be
-\beta F_{\Lambda,\Lambda^\prime}(\beta) \; := \; P_{\Lambda,\Lambda^\prime}(\beta) \; := \; \av{ {\cal P}_{\Lambda,\Lambda^\prime}(\beta) }\; .
\ee
\item $R$-{\it product random Boltzmann-Gibbs state}.
\be
\Omega_{\Lambda} (-) \; := \;
\sum_{\sigma^{(1)},...,\sigma^{(R)}}(-)\,
\frac{
e^{-\beta[H_\Lambda(\s^{(1)})+\cdots
+H_\Lambda(\sigma^{(R)})]}}{[{\cal Z}_{\Lambda}(\beta)]^R}
\; .
\label{omega}
\ee
\item {\it Quenched equilibrium state}.
\be
\<-\>_{\Lambda} \, := \av{\Omega_{\Lambda} (-)} \; .
\ee
\item\label{obs} {\it Observables}.\\
For any smooth bounded function $G(c_{\Lambda})$
(without loss of generality we consider $|G|\le 1$ and no assumption of
permutation invariance on $G$ is made) of the covariance matrix
entries we introduce the random (with respect to $\<-\>$)
$R\times R$ matrix of elements $\{q_{k,l}\}$ (called {\it generalized
overlap}) by the formula
\be
\<G(q)\> \; := \; \av{\Omega (G(c_{\Lambda}))} \; .
\ee
E.g.:
$G(c_\Lambda)= c_{\Lambda}(\sigma^{(1)},\sigma^{(2)})c_{\Lambda}(\sigma^{(2)}
,\sigma^{(3)})$
\be
\<q_{1,2}q_{2,3}\> \; = \;
\av{\sum_{\sigma^{(1)},\sigma^{(2)},\sigma^{(3)}}
c_{\Lambda}(\sigma^{(1)},\sigma^{(2)})c_{\Lambda}(\sigma^{(2)},\sigma^{(3)})
\;\frac{
e^{-\beta[\sum_{i=1}^{3}H_\Lambda(\s^{(i)})]}}{[{\cal Z}(\beta)]^3}}
\ee
\end{enumerate}
{\it Remark: In the following, whenever there is no risk of confusion,
the volume dependency in the quenched state or in the thermodynamic quantities
will be dropped.}
\section{Preliminary: interpolation on the circle}\label{sec_interp}
Let $\xi = \{\xi_i\}_{1\leq i\leq n}$ and $\eta = \{\eta_i\}_{1\leq i\leq n}$ be two independent
families of centered Gaussian random variables, each having covariance matrix $\CC$, i.e.
\beq
\Av(\xi_i\xi_j)   & = & \CC_{i,j} \nonumber\\
\Av(\eta_i\eta_j) & = & \CC_{i,j} \nonumber\\
\Av(\xi_i\eta_j)  & = & 0.
\eeq
Consider the following linear combination of $\xi$ and $\eta$
\beq\label{interp_cerchio}
x_i(t) & = & f(t) \xi_i + g(t)\eta_i \nonumber
\eeq
where the parameter $t\in [a,b]\subset \R$ and the two functions $f(t),g(t)$  take real values
subject to the constraint
\be
\label{cerchio}
f(t)^2+g(t)^2=1\;.
\ee
Chosing $f(t)=\cos (t)$, $g(t)=\sin (t)$ we obtain:
\beq
x_i(t) =  \cos(t)\, \xi_i +\sin(t)\, \eta_i \nonumber.
\eeq
Because of the constraint (\ref{cerchio}), for any given time $t\in[a,b]$, the new centered Gaussian family
$x(t)=\{x_i(t)\}_{1\leq i\leq n}$ has the same covariance structure of $\xi$  and $\eta$:
\beq \label{cov-tt}
\Av(x_i(t)x_j(t))  & = & \CC_{i,j},\nonumber
\eeq
and hence the same distribution, independently of $t$ (i.e. $x(t)$ is a stationary Gaussian process).

In the abstract set-up described above, we regard $x(t)$ as an interpolating Hamiltonian which
is a linear combination of the random Hamiltonians $\xi$ and $\eta$, with $t$-dependent weights
that are the coordinates of a point on the circle of unit radius.\footnote{It is probably worth noting that
any other parametrization of the unit circle would lead to the same expression as in (\ref{varX1}).} We introduce the interpolating random pressure
\footnote{Here, in defining the interpolating (random) pressure, we absorb the temperature in the Hamiltonian.}:
\be
{\cal P}(t) = \ln Z(t)=  \ln\sum_{i=1}^n e^{x_i(t)}\;,
\ee
and the notation $\<C_{1,2}\>_{t,s}$ to denote the expectation of the
covariance matrix in the deformed quenched state constructed from two
independent copies with Boltzmann weights $x(t)$, respectively $x(s)$.
Namely:
\beq
\<C_{1,2}\>_{t,s}=\Av\sum_{i,j=1}^n
\CC_{i,j} \frac {e^{x_i(t)+x_j(s)}}{Z(t)Z(s)}\;.
\eeq
The definition is extended in the obvious way to more than two copies.
We will be interested in the random variable given by the difference of the pressures evaluated
at the boundaries values
\be
\X(a,b) = {\cal P}(b)-{\cal P}(a)\;.
\ee
The following lemma gives an explicit expression for the first two moments
of this random variable.
\begin{lemma}
\label{lemma1}
For the random variable $\X(a,b)$ defined above we have
\be\label{vaXzero}
\Av(\X(a,b))=0
\ee
and
\beq\label{varX1}
\Av[(\X(a,b))^2] & = &
\int_{a}^{b}  \int_{a}^{b}  dt\;  d s\; k_1(t,s)\<C_{1,2}\>_{t,s}
\\
&-& \int_{a}^{b}  \int_{a}^{b} dt\; d s\;
k_2(t,s) \left [ \<C_{1,2}^2\>_{t,s} -
2\<C_{1,2}C_{2,3}\>_{s,t,s} +\<C_{1,2}C_{3,4}\>_{t,s,s,t}\right]
\nonumber \eeq with \beq
k_1(t,s)=\cos(t-s),\qquad\qquad
k_2(t,s)=\sin^2(t-s). \eeq
\end{lemma}
{\bf Proof.}
The stationarity of the Gaussian process $x(t)$ implies that $\Av ({\cal P}(t))$
is independent of $t$, this proves (\ref{vaXzero}).
As far as the computation of the second moment is concerned, starting from
\beq
\Av(\X(a,b))  =  \int_a^b dt\Av({\cal P}'(t))  =  \int_a^b dt \sum_{i=1}^n \Av\left(x'_i(t)
\frac{e^{x_i(t)}}{Z(t)}\right)
\eeq
we have
\beq
\label{var}
\Av[(\X(a,b))^2] & = & \int_a^b dt \int_a^b ds \Av({\cal P}'(t){\cal P}'(s)) \nonumber \\
& = & \int_a^b dt \int_a^b ds \sum_{i,j=1}^n \Av\left(x'_i(t)x'_j(s) \frac{e^{x_i(t)+x_j(s)}}{Z(t)Z(s)}\right).
\eeq
The computation of the average in the rightmost term of the previous formula, which is reported in Appendix 1,
gives
\beq\label{brontolo}
& &\Av\left(x'_i(t)x'_j(s) \frac{e^{x_i(t)+x_j(s)}}{Z(t)Z(s)}\right)=
\cos(t-s) \langle C_{1,2}\rangle_{t,s}\nonumber\\
&-&\sin^2(s-t)\left( <C_{12}^2>_{t,s}-2<C_{12}C_{23}>_{t,s,t}+<C_{12}C_{34}>_{t,s,s,t}\right)
\eeq
proving (\ref{varX1}).
\qed
\section{Bound on the fluctuations of the free energy difference}
It is a well established fact that the random free energy per particles of Gaussian spin glasses
satisfies concentration inequalities, implying in particular self-averaging.
Here we prove that the same result holds for the variation in the random free energy
(or equivalently the random pressure)
\be
\bll={\cal P}_\Lambda- {\cal P}_{\Lambda,\Lp}
\ee
induced by the change of the signs of the interaction in the subset $\Lp\subseteq \L$.
In general, the fact that the random free energy per particle concentrates around its mean
as the system volume increases  of the free energy can be obtained either by martingales
arguments \cite{PS,CGi2} or by general Gaussian concentration of measure \cite{T,GT2}.
Here we follow the second approach. Our formulation applies to both
mean field and finite dimensional models and, for instance, includes
the non summable interactions in finite dimensions \cite{KS} and the
$p$-spin mean field model as well as the REM and GREM  models.
\par\noindent
Before stating the result, it is useful to notice that, as a consequence of the symmetry
of the Gaussian distribution, the variation of the random pressure has a zero average:
\be\label{av_dif_fre_en}
{\rm Av}(\bll)=0\; .
\ee
\begin{lemma}
\label{martin}
For every subset $\Lp\subset \L$ the disorder fluctuation of the free energy variation $\bll$ satisfies the following inequality: for all $x > 0$
\be
\label{concentra}
\p\,\left(|\bll| \ge x\right) \;\le\; 2 \exp{\left(-\frac{x^2}{8 \pi \b^2 \c |\Lp|}\right)}
\ee
with $\c$ the constant in the thermodynamic stability condition (cfr. Eq. (\ref{thst})).
The variance of the free energy variation satisfies the bound
\be\label{sav}
Var({\bll}) \; = \;  \av{\bll^{2}} \;\le\; 16\, \pi\,  \c\, \beta^2\, {|\Lp|}
\ee
\end{lemma}
{\bf Proof.}
Consider an $s > 0$ and let $x>0$. By Markov inequality, one has
\bea
\label{markov}
\p\,\left\{ \bll \ge x\right\}
& = &
\p\,\left\{\exp [s \bll ] \ge \exp (sx)\right\}
\nonumber\\
&\le & \av{\exp[s\bll]} \; \exp(-sx)
\eea
To bound the generating function
\be
\av{\exp[s \bll]}
\ee
one introduces, for a parameter $t\in [0,\frac \pi 2]$, the following interpolating partition functions:
\bea
Z^+(t)=\sum_{\s  \in \,\Sigma_\Lambda} e^{-\b \cos t\,  H^{(1)}_{\Lp}(\s)-\b H^{(3)}_{\Lambda \setminus \Lp}(\s)-\b \sin t\, H^{(2)}_{\Lp}(\s)},\\
Z^-(t)=\sum_{\s  \in \,\Sigma_\Lambda} e^{\b \cos t\, H^{(1)}_{\Lp}(\s)-\b H^{(3)}_{\Lambda \setminus \Lp}(\s)+\b \sin t\, H^{(2)}_{\Lp}(\s)}\;.
\eea
Here the hamiltonians  $H^{(1)}_{\Lp}(\s)$,  $H^{(2)}_{\Lp}(\s)$, $H^{(3)}_{\Lambda \setminus \Lp}(\s)$, defined according to (\ref{hamiloc}),
depend on three independent copies $\{J^{(1)}_X\}_{X\subset \L},\; \{J^{(2)}_X\}_{X\subset \L},\; \{J^{(3)}_X\}_{X\subset \L}$
of the Gaussian disorder characterized by (\ref{mean_disorder}),(\ref{var_disorder}). Now we are ready to consider the
interpolating function
\be
\phi(t) = \ln  Av_3 Av_1\left\{\exp\left(s \; Av_2\left\{\ln \frac{{Z^+}(t)}{{Z^-}(t)} \right\}\right)\right\}\;,
\ee
where $Av_1\{-\}$,  $Av_2\{-\}$ and  $Av_3\{-\}$ denote expectation with
respect to the three independent families of Gaussian variables $J_X$.
It is immediate to verify that
\be
\phi(0) = \ln {\rm Av} \exp [s \;{\bll}]  \;,
\ee
and, using (\ref{av_dif_fre_en}),
\be
\phi\left (\frac \pi 2\right ) = 0 \;.
\ee
This implies that
\be
\label{bush}
\av{\exp[s\bll]} = e^{\phi(0)-\phi(\frac \pi 2)} = e^{-\int_{0}^{\frac \pi 2} \phi'(t) dt}.
\ee
On the other hand, the function $\phi'(t)$ can be easily bounded. Defining
\be
K(t) = \exp\left (s \; Av_2\left\{\ln \frac{{Z^+}(t)}{{Z^-}(t)} \right\}\right )
\ee
the derivative is given by
\be\
\phi'(t)=\phi'_+(t)+\phi'_-(t)
\ee
where
\be\label{der_phi}
\phi'_+(t)=\frac{s Av_3 Av_1\left\{K(t)Av_2\left\{\frac{{Z^+(t)}^\prime}{Z^+(t)}\right\}\right\}}{Av_3 Av_1 \left\{K(t)\right\}},\quad
\phi'_-(t)=-\frac{s Av_3 Av_1\left\{K(t)Av_2\left\{\frac{{Z^-(t)}^\prime}{Z^-(t)}\right\}\right\}}{Av_3 Av_1 \left\{K(t)\right\}}.
\ee
The first term in the derivative is
\be
\phi'_+(t)=\frac{sAv_3 Av_1\left\{K(t)Av_2\left\{\sum_{\s  \in \,\Sigma_\Lambda} p^+_t(\s)\left[\b\sin t\,H^{(1)}_{\Lp}(\s)-\b\cos t\, H^{(2)}_{\Lp}(\s) \right]\right\}\right\}}
{Av_3 Av_1 \left\{K(t)\right\}}
\ee
where
\be
p^+_t(\s) =\frac{e^{-\b \cos t\,  H^{(1)}_{\Lp}(\s)-\b H^{(3)}_{\Lambda \setminus \Lp}(\s)-\b \sin t\, H^{(2)}_{\Lp}(\s)}}{Z^+(t)}
\ee
Applying the integration by parts formula, a simple computation gives
\bea\label{der_primo_addendo}
& &\b\sin t\,Av_3Av_1\left\{K(t)
\; Av_2\left\{\sum_{\s}  p^+_t(\s)H^{(1)}_{\Lp}(\s) \right \}   \right\}\nonumber\\
& = &
-s \b^2 \sin t\, \cos t\, Av_3Av_1  \left\{K(t) \sum_{X\subset \Lp} \Delta_X^2[s^+_t(X)^2+
s^+_t(X)s^-_t(X)]\right \}
\nonumber\\
&-&
\b^2 \sin t\, \cos t\, Av_3Av_1\left\{K(t)
 \; Av_2\left\{\sum_{\s}{\cal C}_{\Lp}(\s,\s) p^+_t(\s)  \right\}   \right\}
\nonumber\\
& + &
\b^2 \sin t\, \cos t\, Av_3Av_1\left\{K(t)
 \; Av_2\left\{\sum_{\s,\tau}{\cal C}_{\Lp}(\s,\tau) p^+_t(\s)p^+_t(\tau) )   \right\}   \right\}
\eea
and
\bea\label{der_secondo_addendo}
& &-\b \cos t\, Av_3Av_1\left\{K(t)
 \; Av_2\left\{\sum_{\s} p^t(\s) \;H^{(2)}_{\Lp}(\s)  \right\}   \right\}\nonumber\\
&= &
\b^2 \sin t\, \cos t\, Av_3Av_1\left\{K(t)
 \; Av_2\left\{\sum_{\s}{\cal C}_{\Lp} (\s,\s)p^+_t(\s)  \right\}   \right\}
\nonumber\\
& - &
\b^2\sin t\, \cos t\, Av_1\left\{K(t)
 \; Av_2\left\{\sum_{\s,\tau}{\cal C}_{\Lp}(\s,\tau) p^+_t(\s)p^+_t(\tau) )   \right\}   \right\}
\eea
where
\be
s^+_t(X)=Av_2\left\{\sum_{\s}\s_Xp^+_t(\s)\right\},\quad s^-_t(X)=Av_2\left\{\sum_{\s}\s_Xp^-_t(\s)\right\}
\ee
and
\be
p^-_t(\s) =\frac{e^{\b \cos t\,  H^{(1)}_{\Lp}(\s)-\b H^{(3)}_{\Lambda \setminus \Lp}(\s)+\b \sin t\, H^{(2)}_{\Lp}(\s)}}{Z^-(t)}\;.
\ee
Taking the difference between (\ref{der_primo_addendo}) and (\ref{der_secondo_addendo}) one finds that
\be\label{der_phi_primo}
\phi'_+(t)=-s^2 \b^2 \sin t\, \cos t\,\frac{ Av_3Av_1  \left\{K(t) \sum_{X\subset \Lp} \Delta_X^2[s^+_t(X)^2+
s^+_t(X)s^-_t(X)]\right \}}
{Av_3Av_1\{K(t)\}}\;.
\ee
With a similar computation one obtains also
\be\label{der_phi_primo}
\phi'_-(t)=-s^2 \b^2 \sin t\, \cos t\,\frac{ Av_3Av_1  \left\{K(t) \sum_{X\subset \Lp} \Delta_X^2[s^-_t(X)^2+
s^+_t(X)s^-_t(X)]\right \}}
{Av_3Av_1\{K(t)\}}\;,
\ee
then we conclude that
\be
\phi'(t)=-s^2 \b^2 \sin t\, \cos t\,\frac{ Av_3Av_1  \left\{K(t) \sum_{X\subset \Lp} \Delta_X^2[s^+_t(X)+
s^-_t(X)]^2\right \}}
{Av_3Av_1\{K(t)\}}\;.
\ee
Using the thermodynamic stability condition (\ref{pippo}), this yields
\be
|\phi'(t)| \le 4\,\b^2\,\c\,s^2\,|\Lp|
\ee
from which it follows, using (\ref{bush})
\be
\av{\exp[s\bll]} \le \exp\left(2 \pi \b^2\,\c\, s^2\,|\Lp|\right).
\ee
Inserting this bound into the inequality (\ref{markov}) and optimizing over $s$
one finally obtains
\be
\p\,\left(\bll \ge x\right) \;\le\;\exp{\left(-\frac{x^2}{8\pi\,\b^2\,\c\, |\Lp|}\right)}.
\ee
The proof of inequality (\ref{concentra}) is completed by observing
that one can repeat a similar computation for  $\p\,\left(\bll \le -x\right)$.
The result for the variance (\ref{sav}) is then immediately proved, thanks to (\ref{av_dif_fre_en}),
using the identity
\be
\av{\bll^2} = 2 \int_{0}^{\infty} x\; \p(|\bll| \ge x) \;dx\;.
\ee
\qed
\section{Overlap identities from the difference of free energy}
We are now ready to state our first result.
\begin{theorem}
\label{cesare}
Given a volume $\L$, consider the Guassian spin glass
with Hamiltonian (\ref{hami}). For a subvolume $\Lp\subseteq\L$ and
a parameter $t\in [0,\pi]$, let
$$
\omega_t(-) = \sum_{\sigma} (-) e^{-H_\sigma(t)}/Z(t)
$$
with
$$
H_{\sigma}(t) = \cos(t) H^{(1)}_{\Lp}(\s)+ \sin(t) H^{(2)}_{\Lp}(\s)+ \hcl(\s)
$$
be the Boltzmann-Gibbs state which interpolates between the system with Gaussian
disorder and the system with a flipped disorder in the region $\Lp$
( $H^{(1)}_{\Lp}$ and  $H^{(2)}_{\Lp}$ are two independent copies
of the Hamiltonian in the subvolume $\Lp$,
$\hcl(\s)$ is the Hamiltonian in the remaining part of the volume,
they are all independent). Then, the following identities hold
\be
\label{id-freeenergy}
\lim_{\L,\Lp \nearrow \Z^d}
\int_{0}^{\pi}\int_{0}^{\pi} dt\; ds\; \sin^2 (s-t) \left [ \<(c^{\Lp}_{1,2})^2\>_{t,s} - 2\<c^{\Lp}_{1,2}c^{\Lp}_{2,3}\>_{s,t,s}
+\<c^{\Lp}_{1,2}c^{\Lp}_{3,4}\>_{t,s,s,t}\right] = 0
\ee
where $\<(c^{\Lp}_{1,2})^2\>_{t,s}$ (and analogously for the other terms)
is the overlap of region $\Lp\subseteq\L$ in the quenched state
constructed form the interpolating Boltzmann-Gibbs state, e.g.
$$
\<(c^{\Lp}_{1,2})^2\>_{t,s} = \Av(\omega_t\omega_s (c^2_{\Lp}(\s,\tau)))\;.
$$
\end{theorem}

\noindent
{\bf Proof:}
The proof is obtained from a suitable combination
of the results in the previous sections. For a parameter
$t\in [0,\pi]$ we consider the interpolating random pressure
\be
{\cal P}(t) = \ln \sum_{\sigma\in\Sigma_{\L}} e^{x_{\sigma}(t)+ \hcl(\s)}
\ee
where
$$
x_{\sigma}(t) = \cos(t) H^{(1)}_{\Lp}(\s)+ \sin(t) H^{(2)}_{\Lp}(\s)
$$
with $H^{(1)}_{\Lp}(\s), H^{(2)}_{\Lp}(\s)$ two independent copies of
the Hamiltonian for the subvolume $\Lp\subseteq\L$.
The boundaries values give the random pressure of the original system
when $t=0$ and the random pressure of the system with the couplings $J$ flipped
on the subvolume $\Lp$ when $t=\pi$, i.e.
$$
{\cal P}(0) = {\cal P}_{\L}\;,
$$
$$
{\cal P}(\pi) = {\cal P}_{\L,\Lp}\;.
$$
Application of Lemma \ref{lemma1} with $\xi_{\s} = H^{(1)}_{\Lp}(\s)$ and $\eta_{\s} = H^{(2)}_{\Lp}(\s)$
(the presence of the additional term $\hcl(\s)$ in the random interpolating pressure does not change the
result in the Lemma, as far as the quenched state is correctly interpreted) gives
\begin{eqnarray}
Var({\cal P}_{\L} - {\cal P}_{\L,\Lp})
& = &
\Lp \int_{0}^{\pi}\int_{0}^{\pi}  dt\;  ds\; \cos(s-t) \<c^{\Lp}_{1,2}\>_{t,s}
\\
&+& ({\Lp})^{2}\int_{0}^{\pi}\int_{0}^{\pi} dt\; ds\; \sin^2 (s-t) \left [ \<(c^{\Lp}_{1,2})^2\>_{t,s} - 2\<c^{\Lp}_{1,2}c^{\Lp}_{2,3}\>_{s,t,s}
+\<c^{\Lp}_{1,2}c^{\Lp}_{3,4}\>_{t,s,s,t}\right]
\nonumber
\end{eqnarray}
On the other hand, Lemma (\ref{martin}) tell us that $Var({\cal P}_{\L} - {\cal P}_{\L,\Lp})$
is bounded above by a constant times the subvolume $\Lp$. As a consequence,
the statement of the theorem follows.
\qed

{\it Remark:} When expressed in terms of the spin variables
the polynomial in the integral (\ref{id-freeenergy})
involves generalized truncated correlation functions.
Indeed, for the model defined in Section 2,
we have the following expressions
\beq
\omega_{t,s}((C^{\Lp}_{1,2})^2)=\sum_{X,Y\subset \Lp} \Delta_X^2 \Delta_Y^2
\omega_t(\si_X^{(1)}\si_Y^{(1)})\omega_s(\si_X^{(2)}\si_Y^{(2)})\nonumber\\
\omega_{s,t,s}(C^{\Lp}_{1,2}C^{\Lp}_{2,3})=\sum_{X,Y\subset \Lp} \Delta_X^2 \Delta_Y^2
\omega_s(\si_X^{(1)})\omega_t(\si_X^{(2)}\si_Y^{(2)})\omega_s(\si_Y^{(3)})\nonumber\\
\omega_{t,s,s,t}(C^{\Lp}_{1,2}C^{\Lp}_{3,4})=\sum_{X,Y\subset \Lp} \Delta_X^2 \Delta_Y^2
\omega_t(\si_X^{(1)})\omega_s(\si_X^{(2)})\omega_s(\si_Y^{(3)})\omega_t(\si_Y^{(4)})
\eeq
thus
\beq\label{comb_mom_av}
& &\omega_{t,s}((c^{\Lp}_{1,2})^2)-2\;\omega_{s,t,s}(c^{\Lp}_{1,2}c^{\Lp}_{2,3})+\omega_{t,s,s,t}(c^{\Lp}_{1,2}c^{\Lp}_{3,4})=\\
& &\frac{1}{|\Lp|^2}\sum_{X,Y\subset \Lp} \Delta_X^2 \Delta_Y^2 \left [\omega_{t}(\si_X\si_Y)- \omega_{t}(\si_X)\omega_{t}(\si_Y)\right ]
\left [\omega_{s}(\si_X\si_Y)- \omega_{s}(\si_X)\omega_{s}(\si_Y)\right ]\nonumber
\eeq
where replica indices have been dropped.
For the Edwards-Anderson model, which is obtained with
$\Delta_X^2 = 1$ if $X\in B'=\{(n,n^\prime)\in \Lp\times\Lp,|n-n^\prime|=1\}$ and $\Delta_X^2=0$
otherwise, the linear combination (\ref{comb_mom_av}) of the moments of the link-overlap in the region $\Lp$
is written in terms of truncated correlation functions, that is
\beq
& &\omega_{t,s}((c^{\Lp}_{1,2})^2)-2\;\omega_{s,t,s}(c^{\Lp}_{1,2}c^{\Lp}_{2,3})+\omega_{t,s,s,t}(c^{\Lp}_{1,2}c^{\Lp}_{3,4})=\nonumber\\
& &\frac{1}{|\Lp|^2}\sum_{b,\bp\in B'}\left [\omega_{t}(\si_b\si_\bp)- \omega_{t}(\si_b)\omega_{t}(\si_\bp)\right ]
\left [\omega_{s}(\si_b\si_\bp)- \omega_{s}(\si_b)\omega_{s}(\si_\bp)\right ],\label{unodueunochi}
\eeq
with $\si_b=\si_n\si_n^\prime$, if $b=(n,n^\prime)\in B'$.
\section{Overlap identities from the difference of internal energy}

In this section we study the change in the internal energy after a
flip of the couplings. We consider only the case of the flip of all
the couplings in the entire volume.

Let us consider two centered gaussian families $\xi=\{\xi_i\}_{1\le
i\le n}$, $\eta=\{\eta_i\}_{1\le i\le n}$ with covariance structure
given by
\beq \av{\xi_i\xi_j}=\av{\eta_i\eta_j}={\cal C}_{i,j}
\eeq
with ${\cal C}_{i,i}=N$. We assume the thermodynamic stability
condition to hold. It follows that $N$ is proportional to the
volume. For example, in the case of the Edwards-Anderson model on a
$d$-dimensional lattice we would have $N=d|\Lambda|$. We introduce
the random free energies
\beq {\cal P}_\xi(\beta)=\ln
Z_\xi(\beta)=\ln \sum_i e^{-\beta \xi_i},\quad {\cal
P}_\eta(\beta)=\ln Z_\eta(\beta)=\ln \sum_i e^{-\beta \eta_i},
\eeq
with the random Boltzmann-Gibbs state $\bgh(-),\bgk(-)$  and their
quenched versions:
\beq \<-\>_{\xi}=\avh \bgh(-),\qquad
\<-\>_{\eta}=\avk \bgk(-).
\eeq
With a slight abuse of notation we
will use the previous symbols also to denote  the product state
acting on the replicated system. The free energy difference,
obtained flipping the hamiltonian $\eta$,
\beq \X(\beta)={\cal
P}_\xi(\beta)-{\cal P}_{-\eta}(\beta)\equiv \ln \sum_i e^{-\beta
\xi_i} - \ln \sum_i e^{\beta \eta_i},
\eeq
has a $\beta$-derivative
given by the difference between the internal energies: \beq
\X^\prime(\beta)=-\bgh(\xi)-\bgmk(\eta). \eeq Using the symmetry of
the distribution of $\eta$, we have the identities\footnote{ Indeed,
from the symmetry of the gaussian distribution, we have that for any
function $f(\eta)$ the following equalities hold:
$\avk f(\eta)=\avk f(-\eta)=\avmk f(-\eta)$.
In particular if $g$ is a function of the configurations of the replicated system, applying the previous remark
to $f(\eta)=\bgk(g)$ we obtain:
$\<g\>_{\eta}\equiv \avk \bgk (g)= \avk \bgmk (g)= \avmk \bgmk (g)\equiv \<g\>_{-\eta}$.

These properties will be tacitly used  several time in this section.}
\beq
&&\avh \bgh(\xi)=-\beta(N-\avh \bgh (C_{1,2}))=-\beta(N- \<C_{1,2}\>_\xi)\label{en_int_h}\\
&&\avk \bgmk(\eta)=\beta(N-\avk \bgmk (C_{1,2}))=\beta(N-
\<C_{1,2}\>_\eta).\label{en_int_k}
\eeq
The above formulae show that
the disorder average of $\X^\prime(\beta)$ vanishes
\beq
\avhk(\X^\prime(\beta)) =\beta (\<C_{1,2}\>_{\eta}-
\<C_{1,2}\>_{\xi})=0,
\eeq
since, obviously, $\<C_{1,2}\>_{\eta} =
\<C_{1,2}\>_{\xi}$. Here $C_{1,2}=\{\mathcal{C}_{i,j}\}_{i,j}$
represents the covariance matrix whose entries are regarded as
configurations of two replicas labeled 1 and 2. Thus, using the
identity $\avk(\bgmk(\eta)^2)=\avh(\bgh(\xi)^2)$, we have that the
variance of $\X^\prime(\beta)$ is given by:
\beq\label{avhkxprime}
\avhk(\X^\prime(\beta)^2)=2\avh(\bgh(\xi)^2)+2\avhk(\bgh(\xi)\bgmk(\eta)).
\eeq
Using the integration by parts formula, we obtain that
\beq\label{av_enintquad}
\avh(\bgh(\xi)^2)=
\avh\sum_{i,j} C_{i,j}\frac{e^{-\beta \xi_i-\beta \xi_j}}{Z_\xi(\beta)^2}
+\avh\sum_{i,j}\sum_{k,\ell} C_{i,k} C_{j,\ell} \frac{\partial^2
}{\partial \xi_\ell \partial \xi_k}\left [ \frac{e^{-\beta \xi_i-\beta \xi_j}}{Z_\xi(\beta)^2} \right ].
\eeq
The second term in the right-hand side of the previous formula requires a repeated application of the integration
by parts formula, which gives:
\beq
\avh\sum_{i,j}\sum_{k,\ell} C_{i,k} C_{j,\ell} \frac{\partial^2
}{\partial \xi_\ell \partial \xi_k}\left [ \frac{e^{-\beta \xi_i-\beta \xi_j}}{Z_\xi(\beta)^2}\right ]
 = \beta^2N(N-2\<C_{1,2}\>_{\xi})+\beta^2\<C_{1,2}^2\>_{\xi}\nonumber\\
-6\beta^2\<C_{1,2}C_{2,3}\>_{\xi}+6\beta^2\<C_{1,2}C_{3,4}\>_{\xi}.\nonumber
\eeq
Since the first term in the right-side of (\ref{av_enintquad}) is quenched average of
$C_{1,2}$, we conclude that
\beq
\avh(\bgh(\xi)^2)&=&\<C_{1,2}\>+\beta^2N(N-2\<C_{1,2}\>)+\beta^2\<C_{1,2}^2\>\nonumber\\
&-&6\beta^2\<C_{1,2}C_{2,3}\>+6\beta^2\<C_{1,2}C_{3,4}\>\label{primo_addendo}
\eeq
dropping, here and in what follows, the unessential reference
to $\xi$ in the quenched averages. If the two families $\xi$ and
$\eta$ were independent, then in (\ref{avhkxprime}) the average of
the product would factorize $\avhk(\bgh(\xi)\bgmk(\eta))=-\beta^2(N-
\<C_{1,2}\>)^2$ giving:
\beq
\avhk(\X^\prime(\beta)^2)&=&2\<C_{1,2}\>+2\beta^2\left(
\<C_{1,2}^2\>-\<C_{1,2}\>^2\right)+12\beta^2
\left(\<C_{1,2}C_{3,4}\>-\<C_{1,2}C_{2,3}\> \right).\nonumber
\eeq
In this case the self averaging of the normalized quantity
$\X^\prime(\beta)^2/N$ (see Theorem 2) would lead, in the large
volume limit $N\rightarrow \infty$, to the well known identity
\cite{G2}
\beq \langle c_{1,2}c_{2,3}\rangle - \langle
c_{1,2}c_{3,4}\rangle = \frac 1 6 \left( \langle c_{1,2}^2\rangle -
\langle c_{1,2} \rangle^2 \right).
\eeq
However, our concern here is the
computation of  the quadratic fluctuations of $\X^\prime(\beta)$
when the sign of a given hamiltonian $\xi$ is flipped in the whole
volume. Therefore we have to set $\xi$=$\eta$ in (\ref{avhkxprime}).
The computation requires, once again, the repeated use of the
integration by parts formula
\beq\label{avhxprimehmenoh}
\avh(\bgh(\xi)\bgmh(\xi))&=&\avh\left (\sum_{i,j} \xi_i \xi_j
\frac{e^{-\beta \xi_i+\beta
\xi_j}}{Z_\xi(\beta)Z_\xi(-\beta)}\right)
=\avh\sum_{i,j} C_{i,j}\frac{e^{-\beta \xi_i+\beta \xi_j}}{Z_\xi(\beta)Z_\xi(-\beta)} \nonumber \\
&+&\avh \sum_{i,j} \sum_{k,\ell} C_{i,k}  C_{j,\ell}\frac{\partial^2
}{\partial \xi_\ell \partial \xi_k}\left [ \frac{e^{-\beta \xi_i+\beta \xi_j}}{Z_\xi(\beta)Z_\xi(-\beta)}\right ].
\eeq
The average in (\ref{avhxprimehmenoh}) is expressed through a set of {\it mixed} quenched state: for instance, the
first term in right-hand side of the previous equation is
\beq
\langle C_{1,2} \rangle_{+,-}=\avh \sum_{i,j}  \mathcal{C}_{i,j} \frac{e^{-\beta \xi_i+\beta \xi_j}}{Z_{\xi}(\beta)Z_{\xi}(-\beta)}.
\eeq
Generalizing the previous definition we have, for instance, that  $\langle - \rangle_{+,+,-,+}$ represents the thermal
average taken with the usual boltzmannfaktor
(i.e. with the sign $-$ in the exponent) in the first, second and fourth copy, and  with the opposite sign in the third one.
Moreover, the symbol $\langle -\rangle_{+,+,+,\ldots}$,  with all the subscripts $+$ (or $-$, because of the symmetry of the
gaussian distribution),
is the usual quenched measure $\langle\ - \rangle$. The explicit computation gives:
\beq
&&\frac{\partial^2}{\partial \xi_\ell \partial \xi_k}\left [ \frac{e^{-\beta \xi_i+\beta \xi_j}}{Z_\xi(\beta)Z_\xi(-\beta)}\right ]
=-\beta^2N^2+2\beta^2N\langle C_{1,2}\rangle_{+,+}-\beta^2\langle C_{1,2}^2\rangle_{+,-}
+2\beta^2\langle C_{1,2}C_{2,3}\rangle_{+,-,+}\nonumber\\
&&-4\beta^2\langle C_{1,2}C_{2,3} \rangle_{+,+,-} +4\beta^2\langle
C_{1,2}C_{3,4}\rangle_{+,+,+,-} -\beta^2\langle
C_{1,2}C_{3,4}\rangle_{+,+,-,-}-\beta^2\langle
C_{1,2}C_{3,4}\rangle_{+,-,+,-}\nonumber
\eeq
and finally:
\beq\label{varxprimo} &&\avh (\X^\prime(\beta)^2)
=2(\langle C_{1,2}\rangle_{+,+} + \langle C_{1,2}\rangle_{+,-})+2\beta^2 \left(\langle C_{1,2}^2\rangle_{+,+}-\langle C_{1,2}^2\rangle_{+,-} \right )\\
&-&4\beta^2\left ( 3\langle C_{1,2}C_{2,3}\rangle_{+,+,+} - \langle C_{1,2}C_{2,3}\rangle_{+,-,+}+2\langle C_{1,2}C_{2,3}\rangle_{+,+,-}\right)\nonumber\\
&+&2\beta^2\left(6\langle C_{1,2}C_{3,4}\rangle_{+,+,+,+}+4\langle
C_{1,2}C_{3,4}\rangle_{+,+,+,-} -\langle
C_{1,2}C_{3,4}\rangle_{+,+,-,-} -\langle
C_{1,2}C_{3,4}\rangle_{+,-,+,-}\right).\nonumber
\eeq
If we choose
now $\xi$ to be the Hamiltonian family defined in section \ref{def},
we obtain the following:
\begin{theorem}
Consider the Guassian spin glass with Hamiltonian $\xi$ given in (\ref{hami}). In
the infinite volume limit
and for almost all
values of $\beta$, we have
\beq \left [\langle c_{1,2}^2
\rangle_{+,+} -\langle c_{1,2}^2 \rangle_{+,- } \right ]
-2 \left [ 3\langle c_{1,2}c_{2,3} \rangle_{+,+,+}-\langle c_{1,2}c_{2,3} \rangle_{+,-,+}+2\langle c_{1,2}c_{2,3} \rangle_{+,+,-}\right ]\nonumber \\
+\left [ 6\langle c_{1,2}c_{3,4} \rangle_{+,+,+,+}+4\langle
c_{1,2}c_{3,4} \rangle_{+,+,+,-}-\langle c_{1,2}c_{3,4}
\rangle_{+,+,-,-} -\langle c_{1,2}c_{3,4} \rangle_{+,-,+,-}\right
]=0
\eeq
where $\<c_{1,2}^2\>_{+,-}$ (and analogously for the
other terms) is the overlap expectation in the quenched state
constructed form the mixed Boltzmann-Gibbs state with one copy given by
the original system and the other copy given by the flipped systems,
e.g.
$$
\<c_{1,2}^2\>_{+,-} = \Av(\omega_\xi\omega_{-\xi}
(c_{\Lambda}^2(\s,\tau)))\;.
$$
\end{theorem}
{\bf Proof.}\par\noindent The proof is a simple consequence of well
known results. The sequence of convex functions ${\cal
P}_\xi(\beta)/N$ converges almost everywhere in $J$ to the limiting
value $a(\beta)$ of its average and the convergence is self
averaging (i.e. $\mathrm{Var}({\cal P}_\xi(\beta)/N)\rightarrow 0$).
By general convexity arguments \cite{RU}
it follows that the sequence of derivatives ${\cal
P}_\xi^\prime(\beta)/N$ converges to $u(\beta)=a^\prime(\beta)$
almost everywhere in $\beta$ and also that the convergence is self
averaging ($\mathrm{Var}({\cal P}_\xi^\prime(\beta)/N)\rightarrow
0$, $\beta$-a.e.) \cite{S,OTW}. These remarks apply obviously also
to ${\cal P}_{-\xi}(\beta)/N$ and to its derivative, with the same
limiting functions $a(\beta)$ and $a^\prime(\beta)$. Thus we have
that $\X(\beta)/N={\cal P}_\xi(\beta)/N-{\cal P}_{-\xi}(\beta)/N$
and its derivative $\X^\prime(\beta)/N$ vanish a.e. in $J$ in the
large volume limit. Moreover,
$\mathrm{Var}(\X^\prime(\beta)/N)=\mathrm{Var}({\cal
P}_\xi^\prime(\beta)/N)+\mathrm{Var}({\cal
P}_{-\xi}^\prime(\beta)/N) -2\mathrm{cov}\left({\cal
P}_{\xi}^\prime(\beta)/N,{\cal P}_{-\xi}^\prime(\beta)/N\right)$,
thus estimating the covariance with the Cauchy-Schwartz inequality
we have \beq \mathrm{Var}(\X^\prime(\beta)/N)\le 4\mathrm{Var}({\cal
P}_\xi^\prime(\beta)/N)\rightarrow 0,\quad \beta -a.e. \eeq for
$N\rightarrow \infty$. Therefore, dividing (\ref{varxprimo}) by
$N^2$ and taking the limit we obtain the result. \qed

\section{Triviality of the Random Field model}
In this section we compute explicitly the expression appearing in
Theorem \ref{cesare} \beq\label{unodueuno} \<c_{1,2}^2\>_{t,s} -2
\<c_{1,2}c_{2,3}\>_{s,t,s}+\<c_{1,2}c_{3,4}\>_{t,s,s,t} \eeq in the
simple case of the random field. We will show that this linear
combination of overlap moments vanishes pointwise for all values of
$t$ and $s$. We will then deduce the triviality of the order
parameter for the random field model.

We consider two families $J_i$ and $\tilde{J}_i$ for $i=1,\ldots, N$
of independent normally distributed centered random variables with
variance 1:
\beq \Av(J_iJ_j)=\Av(\tilde{J}_i
\tilde{J}_j)=\delta_{i,j},\quad \Av(J_i \tilde{J}_j)=0,
\eeq
and the
random field hamiltonians
\beq \label{nerone}\xi_\sigma=\sum_{i=1}^N
J_i\sigma_i,\quad \eta_\sigma= \sum_{i=1}^N \tilde{J}_i\sigma_i.
\eeq
where $\sigma_i=\pm 1$. We have that $\xi=\{\xi_\sigma\}_\si$ and
$\eta=\{\eta_\sigma\}_\si$ are two independent centered Gaussian families
(each having $n=2^N$ elements indexed by the configurations
$\sigma$, $N$ being the volume) and covariance structure given by:
\beq\label{cov-randomf}
&\Av(\xi_\sigma \xi_\tau)\equiv \CC_{\si,\tau}=Nq(\sigma,\tau),\nonumber\\
&\Av(\eta_\sigma \eta_\tau)\equiv \CC_{\si,\tau}= Nq(\sigma,\tau),\nonumber\\
&\Av(\xi_\sigma \eta_\tau)=0.
\eeq
where $q(\sigma,\tau)$ is the {\it site overlap} of the two configurations $\si$ and $\tau$:
\beq
q(\sigma,\tau)=\frac 1 N \sum_{i,j=1}^N \sigma_i\tau_j\; .
\eeq
The interpolating Hamiltonian:
\beq
x_\si(t)  =  \cos(t) \xi_\si + \sin(t)\eta_\si, 
\eeq
which is a stationary Gaussian process with the same distribution of $\xi$ and $\eta$:
\beq
\Av(x_\sigma(t)x_\tau(t))   =  N q(\sigma,\tau),
\eeq
defines the quenched deformed state on the replicated system, whose averages are
denoted with the usual notation, e.g. $\langle - \rangle_{t,s}$, $\langle - \rangle_{s,t,s}\ \ldots$ .
\begin{theorem}
\label{albano} Consider the random field spin glass with Hamiltonian
(\ref{nerone}). In the limit $N\rightarrow \infty$ and for all
values of $t$ and $s$ we have \beq
\g_1\<q_{1,2}^2\>_{t,s}+\g_2\<q_{1,2}\>_{t,s}^2+\g_3\<q_{1,2}q_{2,3}\>_{s,t,s}+\g_4\<q_{1,2}q_{3,4}\>_{t,s,s,t}=0
\eeq for any choice of real $\g_1,\g_2,\g_3,\g_4$ with
$\g_1+\g_2+\g_3+\g_4=0$.\par\noindent
\end{theorem}
{\bf Proof} The simple proof relies on
the following identities, derived in Appendix 2:
\beq
\<C_{1,2}\>_{t,s}^2&=&\sum_{i=1}^N (\Av\big(\tanh(G_{i}(t))\tanh(G_{i}(s))\big))^2+\mathcal{Q}_N(t,s),\label{av_c12quad}\\
\<C_{1,2}^2\>_{t,s}&=&1+\mathcal{Q}_N(t,s),
\label{av_c12q}\\
\<C_{1,2}C_{2,3}\>_{s,t,s}&=&\sum_{i=1}^N\Av\left(\tanh^2(G_i(s))\right)+\mathcal{Q}_N(t,s),
\label{av_c12c23}\\
\<C_{12}C_{34}\>_{t,s,s,t}&=&\sum_{j=1}^N \Av\left(\tanh^2(G_j(t))\tanh^2(G_j(s))\right)+\mathcal{Q}_N(t,s),
\label{av_c12c_34}
\eeq
where
\beq\label{defGi}
G_i(t)=\cos(t)J_i  + \sin(t) \tilde{J}_i,
\eeq
and $\mathcal{Q}_N(t,s)$ is a term of order $N^2$, see (\ref{defQ}). Thus:
\beq
&&\g_1\<C_{1,2}^2\>_{t,s}+\g_2\<C_{1,2}\>_{t,s}^2+\g_3\<C_{1,2}C_{2,3}\>_{s,t,s}+\g_3\<C_{1,2}C_{3,4}\>_{t,s,s,t}=\nonumber\\
&&\g_1+\g_2\sum_{i=1}^N (\Av\big(\tanh(G_{i}(t))\tanh(G_{i}(s))\big))^2 +\g_3\sum_{i=1}^N\Av\left(\tanh^2(G_i(s))\right)\nonumber\\
&&+\g_4\sum_{j=1}^N \Av\left(\tanh^2(G_j(t))\tanh^2(G_j(s))\right)\nonumber\\
&&+(\g_1+\g_2+\g_3+\g_4)\mathcal{Q}_N(t,s),\nonumber
\eeq
i.e. the linear combination of the covariance matrix moments is of order $N$. Thus, since $|\tanh (x)|<1$, we have
\beq
&&\left | \g_1\<C_{1,2}^2\>_{t,s}+\g_2\<C_{1,2}\>_{t,s}^2+\g_3\<C_{1,2}C_{2,3}\>_{s,t,s}+\g_4\<C_{1,2}C_{3,4}\>_{t,s,s,t} \right |\nonumber\\
&&\le |\g_1|+(|\g_2|+|\g_3|+|\g_4|)N,\nonumber
\eeq
which can be rewritten, using the overlaps $q_{1,2},\; q_{2,3},\; q_{3,4}$ between replicas, as
\beq
&&\left | \g_1\<q_{1,2}^2\>_{t,s}+\g_2\<q_{1,2}\>_{t,s}^2+\g_3\<q_{1,2}q_{2,3}\>_{s,t,s}+\g_4\<q_{1,2}q_{3,4}\>_{t,s,s,t} \right |\nonumber\\
&&\le \frac{|\g_2|+|\g_3|+|\g_4|}{N}+\frac{|\g_1|}{N^2}.
\eeq
\qed
\par\noindent
Among the relations of theorem \ref{albano}, in the thermodynamic limit, we find the identity of
theorem \ref{cesare} for the values $\g_1=1, \g_2=0, \g_3=-2, \g_4=1$:
\beq\label{unodueuno0}
\<q_{1,2}^2\>_{t,s}-2\<q_{1,2}q_{2,3}\>_{s,t,s}+\<q_{1,2}q_{3,4}\>_{t,s,s,t}=0
\eeq
and the Ghirlanda-Guerra identities: for $\g_1=1, \g_2=1, \g_3=-2, \g_4=0$ we find
\beq\label{gg1212}
\<q_{1,2}q_{2,3}\>_{s,t,s}=\frac 1 2
\<q_{1,2}^2\>_{t,s} + \frac 1 2 \<q_{1,2}\>_{t,s}^2\; ;
\eeq
for $\g_1=1, \g_2=2, \g_3=0, \g_4=-3$ we find
\beq\label{gg1323}
\<q_{1,2}q_{3,4}\>_{s,t,s}=\frac 1 3
\<q_{1,2}^2\>_{t,s} + \frac 2 3 \<q_{1,2}\>_{t,s}^2\;. \eeq
Using
(\ref{gg1212}) and (\ref{gg1323}) we can express (\ref{unodueuno0})
as:
\beq \label{roma}
\<q_{1,2}^2\>_{t,s}-2\<q_{1,2}q_{2,3}\>_{s,t,s}+\<q_{1,2}q_{3,4}\>_{t,s,s,t}=\frac{1}{3}(\<q_{1,2}^2\>_{t,s}
- \<q_{1,2}\>_{t,s}^2)
\eeq
The identity derived from the flip of the
coupling thus imply a trivial order parameter distribution. Indeed,
since the identity (\ref{unodueuno0}) is true for every $t$ and $s$
we can choose $t=s=0$ and then the  interpolating states reduce to
the usual quenched Boltzmann-Gibbs state. From Eq. (\ref{roma}) we
deduce a trivial overlap distribution.

\vspace{1.cm} \noindent {\bf Acknowledgements:} We thank S. Graffi
for many discussions over stiffness exponent and interpolation,
G. Parisi for discussions on interface problems on spin glasses and for
suggesting to us the check of the new identity on the
random field model. We thank C.Newman and D.Stein for showing
to us the proof by martingale methods of the property mentioned in the
introduction and stated in \cite{NS}. We also thank Francesco Guerra for 
suggesting improvements in the presentation and Aernout van Enter for
useful comments. The authors acknowledge EURANDOM
for the warm ospitality during the "Workshop on Statistical Mechanics and Applications",
the grants Cultaptation (EU) and Funds for Strategic Research (University of Bologna).

\section{Appendix 1}
In this appendix we will use the Gaussian integration by parts formula for correlated Gaussian random
variables $z_1,\ldots,z_n$:
\beq\label{gauss-per-parti}
\Av(z_j\psi(z_1,\ldots,z_n))=\sum_{i=1}^n \av{z_jz_i}\av{\frac{\partial \psi (z_1,\ldots,z_n)}{\partial z_i}},
\eeq
to compute the second moment of the pressure difference $\mathcal{X}(a,b)$. We have to evaluate the
average inside the integral (\ref{var})
\beq\label{gongolo}
&&\sum_{i,j=1}^n \Av\left(x'_i(t)x'_j(s)B(i,j;t,s) \right)=\\
&=&\sint\sins \sum_{i,j=1}^n \Av \left( \xi_i \xi_j B(i, j;t,s) \right)
-\sint\coss \sum_{i,j=1}^n \Av \left( \xi_i \eta_j B(i,j;t,s) \right)\nonumber\\
&-&\sins\cost  \sum_{i,j=1}^n\Av \left( \xi_j \eta_i B(i,j;t,s) \right)\nonumber
+ \cost\coss \sum_{i,j=1}^n \Av \left( \eta_i \eta_j B(i,j;t,s) \right).\nonumber
\eeq
where, for the sake of notation, we have introduced the symbol
$$
B(i,j;t,s)=\frac{e^{x_i(t)+x_j(s)}}{Z(t)Z(s)}.
$$
Applying (\ref{gauss-per-parti}) twice, we obtain
\beq
\label{t1pp}
\Av \left ( \xi_i \xi_j B(i,j;t,s)\right )=
\cc_{i,j}\Av\left(B(i,j;t,s)\right)+\sum_{k,\ell=1}^n \cc_{i,k}\cc_{j,\ell}
\Av\left ( \frac{\partial^2}{\partial \xi_k \xi_\ell} B(i,j;t,s)\right)
\eeq
\beq
\label{t4pp}
\Av \left ( \eta_i \eta_j B(i,j;t,s)\right )=
\cc_{i,j}\Av\left(B(i,j;t,s)\right)+\sum_{k,\ell=1}^n \cc_{i,k}\cc_{j,\ell}
\Av\left ( \frac{\partial^2}{\partial \eta_\ell \eta_k}B(i,j;t,s)\right)
\eeq
\beq
\label{t2pp}
\Av \left ( \xi_i \eta_j B(i,j;t,s)\right )=
\Av \left ( \xi_j \eta_i B(i,j;t,s)\right )=
\sum_{k,\ell=1}^n \cc_{i,k}\cc_{j,\ell}
\Av\left ( \frac{\partial^2}{\partial \xi_k \eta_\ell} B(i,j;t,s)\right)
\eeq
The combination of the first two terms in the right hand sides of (\ref{t1pp}) and (\ref{t4pp}) with the trigonometric
coefficients given by (\ref{gongolo}) produce the quenched expectation  $\cos(t-s) \langle C_{1,2}\rangle_{t,s}$.\par\noindent
The explicit computation of the derivatives is long but not difficult; the result is:
\beq
&\frac{\ds\partial^2}{\ds\partial \xi_k \partial \xi_\ell}B(i,j;t,s)=B(i,j;t,s)\left\{\costq A_1+\cossq A_2 +\cost\coss(A_3+A_4)\right\},\nonumber\\
&\frac{\ds\partial^2}{\ds\partial \eta_k \partial \eta_\ell}B(i,j;t,s)=B(i,j;t,s)\left\{\sinsq A_1+\sinsq A_2 +\sint\sins(A_3+A_4)\right\},\nonumber\\
&\frac{\ds\partial^2}{\ds\partial \xi_k \eta_\ell}B(i,j;t,s)=B(i,j;t,s)\left\{\sint\cost A_1 +\sins\coss A_2\right.\nonumber\\
& \left. + \sint\coss A_3 +\sins\cost A_4\right\},\nonumber
\eeq
where $A_1,A_2,A_3,A_4$ are combinations of Kronecker delta functions depending on the indices $i, j,\ell,k$ and Boltzmann
weights for the hamiltonians $x(t)$ and $x(s)$.\par\noindent
Using the previous formulas for the second derivatives and formulas (\ref{t1pp}),(\ref{t4pp}) and (\ref{t2pp}),  we see that the right hand side of (\ref{gongolo})
contains a linear combination of functions $A_j$ with trigonometric coefficients given by the product of four factors
taken from $\{\cost,\sint,\coss,\sins\}$. It is not difficult to recognize that the coefficient of $A_3$ is $-\sin^2(s-t)$
while the other are zero. Thus,
\beq
\sum_{i,j=1}^n \Av\left(x'_i(t)x'_j(s)B(i,j;t,s) \right)=\cos(t-s) \langle C_{1,2}\rangle_{t,s}
-\sin^2(s-t)\mbox{Av}\sum_{i,j=1}^n\sum_{k,\ell=1}^n
{\cc_{i,k}\cc_{j,\ell}}A_3B(i,j;t,s)\nonumber
\eeq
and since
$$
A_3=\dd{\ell}{i}\dd{k}{j}-\dd{\ell}{i}\frac{e^{x_k(s)}}{Z(s)}-\dd{k}{j}\frac{e^{x_\ell(t)}}{Z(t)}+
\frac{e^{x_\ell(t)}}{Z(t)}\frac{e^{x_k(s)}}{Z(s)}
$$
we obtain
$$
\mbox{Av}\sum_{i,j=1}^n\sum_{k,\ell=1}^n
{\cc_{i,k}\cc_{j,\ell}}A_3B(i,j;t,s)=<C_{12}^2>_{t,s}-2<C_{12}C_{23}>_{t,s,t}+<C_{12}C_{34}>_{t,s,s,t}
$$
which proves (\ref{brontolo}).
\section{Appendix 2}
In this appendix we prove the identities (\ref{av_c12quad}),(\ref{av_c12q}),(\ref{av_c12c23}),(\ref{av_c12c_34}).
Recalling the definition (\ref{defGi}) of the Gaussian variables $G_i(t)$, we can define the interpolating partition function
\beq
&Z(t)=\sum_\sigma \exp(x_\si(t))=\sum_\sigma \exp(\sum_{i=1}^N G_i(t)\sigma_i).
\eeq
A simple computation shows that
\beq
Z(t)=2^N \prod_{i=1}^N \cosh G_i(t).
\eeq
For any integer $M<N$,  we consider the sublattice $\Lambda_M=\{N-M+1,\ldots, N\}\subseteq \Lambda_N\equiv\{1,\ldots,N\}$
with its spin-configuration space $S_M=\left \{-1,1\right \}^{\Lambda_M}$ and the subspace
$S_M^+=\{(+1,\sigma_{N-M+2},\ldots,\sigma_N), \sigma_i=\pm 1\}$ (we will drop the subscript when it is equal to $N$, e.g. $S\equiv S_N$, $S^+\equiv S_N^+$, etc.).
The  interpolating Boltzmann-Gibbs random state on the lattice $\Lambda_M$ is
\beq
& &\omega_{t}^M(f)=\frac{\sum_{\sigma \in S_M} f(\si)\exp\left(\sum_{i=N-M+1}^N G_i(t)\sigma_i\right)}{Z_M(t)},\nonumber\\
& & Z_M(t)=\sum_{\sigma\in S_M} \exp\left(\sum_{i=N-M+1}^N G_i(t)\sigma_i\right),
\eeq
where $f$ is a function on $S_M$. This definition extends in the obvious way to the $R$-fold product; for instance
the 2-product measure for the parameter values $t$ and $s$ is given by
\beq
\omega_{t,s}^M(f)=\frac{\sum_{\sigma,\tau\in S_M}f(\sigma,\tau) \exp\left(\sum_{i=N-M+1}^N G_i(t)\sigma_i+\sum_{i=N-M+1}^N G_i(s)\tau_i\right)}{Z_M(t)Z_M(s)}
\eeq
where $f$ is a function on $S_M\times S_M$.
In the sequel we will also write $\omega_{t,s}(-)$ instead of $\omega_{t,s}^N(-)$.\par\noindent
The computation of the moments of the covariance matrix is done evaluating
(by induction on $M$) the averages of the products of the overlaps between configurations of $S_M$
\beq
q_M(\sigma,\tau)=\frac{1}{M}\sum_{i=N-M+1}^N \sigma_i\tau_i\;.
\eeq
Indeed, the explicit computation shows that:
\beq\label{avn_avnmeno1}
\omega_{t,s}^N(q_N)=\frac{1}{N}\tanh G_1(t)\tanh G_1(s) + \frac{N-1}{N}\omega_{t,s}^{N-1}(q_{N-1}),
\eeq
then iterating the previous formula $N-1$ times we obtain
\beq
\omega^N_{t,s}(q_N)=\frac 1 N \sum_{j=1}^{N}\tanh(G_j(t))\tanh(G_j(s)).
\label{avn}
\eeq
since $\omega^1_{t,s}(q_1)=\tanh(G_N(t))\tanh(G_N(s))$. Recalling the relation between overlaps and covariance (\ref{cov-randomf})
and taking the average with respect to the disorder, we obtain:
\beq
\<C_{1,2}\>_{t,s}=\sum_{i=1}^N \Av\big(\tanh(G_{i}(t))\tanh(G_{i}(s))\big),
\eeq
thus
$$
\<C_{1,2}\>_{t,s}^2=\sum_{i=1}^N (\Av\big(\tanh(G_{i}(t))\tanh(G_{i}(s))\big))^2+\mathcal{Q}_N(t,s)
$$
where
\beq\label{defQ}
\mathcal{Q}_N(t,s)=2\sum_{1\le j<\ell \le N}\Av\left[ \tanh(G_j(t))\tanh(G_j(s))\right]\Av \left [\tanh(G_\ell(t))\tanh(G_\ell(s))\right]
\eeq
is a term of order $N^2$. This proves (\ref{av_c12quad}).\par\noindent
For the squared overlap the following relation holds
\beq
\omega_{t,s}(q_N^2)=\frac{1}{N}+\frac{2}{N^2}\sum_{j=1}^{N-1}(N-j)\tanh(G_j(t))\tanh(G_j(s))\omega_{t,s}^{N-j}(q_{N-1}).
\eeq
Since for $M\le N$
\beq
\omega^M_{t,s}(q_M)=\frac 1 M \sum_{j=N-M+1}^N\tanh(G_j(t))\tanh(G_j(s))
\label{av_qM}
\eeq
we can write
\beq
\omega_{t,s}(q_N^2)=\frac{1}{N^2}+\frac{2}{N^2}\sum_{1\le j<\ell \le N}\tanh(G_j(t))\tanh(G_j(s))\tanh(G_\ell(t))\tanh(G_\ell(s))
\label{boltzAv_c12q}
\eeq
and finally
$$
\<C_{1,2}^2\>_{t,s}=1+2\sum_{1\le j<\ell \le N}\Av\left[ \tanh(G_j(t))\tanh(G_j(s))\tanh(G_\ell(t))\tanh(G_\ell(s))\right].
$$
From the independence of the random variables $G_i(t)$ (see (\ref{defGi})), we have that the average in the right hand side
of the previous formula factorizes, thus we obtain (\ref{av_c12q}):
$$
\<C_{1,2}^2\>_{t,s}=1+\mathcal{Q}_N(t,s).
$$
\par\noindent
The second term in (\ref{unodueuno}) is computed considering the average of $q_N(\sigma,\gamma)q_N(\gamma,\tau)$ where $\gamma,\sigma,\tau\in S$.
We have
\beq
&\omega^N _{s,t,s}(q_N(\sigma,\gamma)q_N(\gamma,\tau))=\frac{\ds 1}{\ds N^2}\tanh^2(G_1(s))\nonumber\\
&+\frac{\ds 2}{\ds N^2}
\tanh(G_1(t))\tanh(G_1(s))\sum_{j=2}^N\tanh(G_j(t))\tanh(G_j(s))\nonumber\\
&+\left(\frac{\ds N-1}{\ds N}\right)^2\omega_{s,t,s}^{N-1}(q_{N-1}(\sigmap,\gammap)q_{N-1}(\gammap,\taup)).
\eeq
where $\sigmap,\gammap,\taup$ are the restriction of $\sigma,\gamma,\tau$ to $S_{N-1}$. As in the previous cases, iterating
this formula and taking into account that $\omega^1_{s,t,s}(q_1(\sigma,\gamma)q_1(\gamma,\tau))=\tanh^2(G_N(s))$ we obtain
\beq
&\omega^N _{s,t,s}(q_N(\sigma,\gamma)q_N(\gamma,\tau))=\frac{\ds 1}{\ds N^2}\sum_{i=1}^N\tanh^2(G_i(s))\nonumber\\
&+\frac{\ds 2}{\ds N^2}\sum_{j=1}^{N-1}
\tanh(G_j(t))\tanh(G_j(s))\sum_{\ell=j+1}^N\tanh(G_\ell(t))\tanh(G_\ell(s)),
\label{boltzAv_c12c23}
\eeq
then
$$
\<C_{1,2}C_{2,3}\>_{s,t,s}=\sum_{i=1}^N\Av\left(\tanh^2(G_i(s))\right)+\mathcal{Q}_N(t,s).
$$
which proves (\ref{av_c12c23}).
The computation of the last term in (\ref{unodueuno}) is simple because in this case the random product
measure factorizes:
\beq
\omega_{t,s,s,t}(q(\sigma,\tau)q(\gamma,\kappa))=\omega_{t,s}(q(\sigma,\tau))\omega_{s,t}(q(\gamma,\kappa)).
\eeq
Then, using (\ref{avn}) we have
\beq
\omega_{t,s,s,t}(q(\sigma,\tau)q(\gamma,\kappa))=\frac{1}{N^2}\sum_{i,j=1}^N \tanh(G_j(t))\tanh(G_j(s))\tanh(G_i(t))\tanh(G_i(s))
\eeq
and
\beq
\<C_{12}C_{34}\>_{t,s,s,t}=\sum_{i,j=1}^N \Av\left(\tanh(G_j(t))\tanh(G_j(s))\tanh(G_i(t))\tanh(G_i(s))\right)
\eeq
which, using the symmetry of $a_{i,j}=\Av\left(\tanh(G_j(t))\tanh(G_j(s))\tanh(G_i(t))\tanh(G_i(s))\right)$,
gives (\ref{av_c12c_34}):
$$
\<C_{12}C_{34}\>_{t,s,s,t}=\sum_{j=1}^N \Av\left(\tanh^2(G_j(t))\tanh^2(G_j(s))\right)+\mathcal{Q}_N(t,s).
$$


\end{document}